# Detection of [C I] Emission in Nebular Spectra of a Peculiar Type Ia Supernova 2022pul

Jialian Liu[1], Xiaofeng Wang[1,2], Yi Yang[1,3], Alexei V. Filippenko[3], Thomas G. Brink[3], WeiKang Zheng[3],  
Jujia Zhang[4,5,6], Gaici Li[1], and Shengyu Yan[1]  
[1] Physics Department, Tsinghua University, Beijing 100084, People's Republic of China; liu-jl22@mails.tsinghua.edu.cn, wang_xf@mail.tsinghua.edu.cn  
[2] Purple Mountain Observatory, Chinese Academy of Sciences, Nanjing 210023, People's Republic of China  
[3] Department of Astronomy, University of California, Berkeley, CA 94720-3411, USA  
[4] Yunnan Observatories, Chinese Academy of Sciences, Kunming 650216, People's Republic of China  
[5] International Centre of Supernovae, Yunnan Key Laboratory, Kunming 650216, People's Republic of China  
[6] Key Laboratory for the Structure and Evolution of Celestial Objects, Chinese Academy of Sciences, Kunming 650216, People's Republic of China  
Received 2025 January 20; revised 2025 February 21; accepted 2025 February 25; published 2025 March 20

## Abstract

SN 2022pul gains special attention due to its possible origin as a super-Chandrasekhar-mass white dwarf (WD) explosion (also called a 03fg-like Type Ia supernova), which shows prominent [O I], [Ne I], and [Ca II] lines in its late-time spectra taken at ∼+300 days after the time of peak brightness. In this Letter, we present new optical observations of this peculiar object, extending up to over 500 days after peak brightness. In particular, in the $t \sim +515$ days spectrum, we identified for the first time the presence of narrow emission from [C I] $\lambda\lambda 9824, 9850$, which appears asymmetric and quite similar to the accompanied [O I] $\lambda 6300$ line in strength and profile. Based on the violent merger model that accounts well for previous observations but leaves little carbon in the center of the ejecta, this carbon line can be reproduced by increasing the degree of clumping in the ejecta and setting the carbon mass the same as that of oxygen ($\sim 0.06\,M_\odot$) in the innermost region ($\lesssim 2000\,\mathrm{km\,s^{-1}}$). In principle, the central carbon could come from the secondary WD if it is ignited when hit by the shock wave of the explosion of the primary WD and explodes as a Ca-rich supernova, whereas pure deflagration of a super-Chandrasekhar-mass WD can account for such unburnt carbon more naturally.

*Unified Astronomy Thesaurus concepts:* Supernovae (1668); Type Ia supernovae (1728); White dwarf stars (1799)

*Materials only available in the* online version of record: machine-readable tables

## 1. Introduction

Type Ia supernovae (SNe Ia; see, e.g., A. V. Filippenko 1997 for a review of supernova classifications) are widely believed to arise from thermonuclear explosions of carbon–oxygen (CO) white dwarfs (WDs; K. Nomoto et al. 1997; W. Hillebrandt & J. C. Niemeyer 2000). They are important distance indicators owing to the high peak luminosity and width–luminosity relation of their light curves ("Lira–Phillips relation"; M. M. Phillips 1993; M. M. Phillips et al. 1999), which enabled the discovery of the accelerating expansion of the Universe (A. G. Riess et al. 1998; S. Perlmutter et al. 1999). However, the mechanism that triggers the explosion and drives the propagation of the burning front, together with the evolutionary channel of the progenitor system, is still under debate (A. J. Ruiter & I. R. Seitenzahl 2024).

Nebular-phase spectra (i.e., at $t \gtrsim 200$ days after the explosion) carry important information to help determine the explosion mechanism. At such late times, the innermost regions of the ejecta are revealed as the opacity decreases with expansion of the ejecta. Hence, the late-time spectra can be used to examine the abundance distribution of central elements, which varies significantly among different models. For example, a pure deflagration model predicts the presence of unburnt carbon in the central region (C. Kozma et al. 2005; M. Fink et al. 2014); oxygen and neon are centrally produced by burning a secondary WD in violent merger models where the primary WD explodes during the merger with the secondary WD (R. Pakmor et al. 2012).

Supernovae (SNe) similar to SN 2003fg (the prototype of super-Chandrasekhar-mass explosions of WDs; D. A. Howell et al. 2006) are a peculiar subtype of SNe Ia with high luminosities and slowly evolving light curves, which indicate an ejecta mass larger than the Chandrasekhar-mass ($M_{\mathrm{Ch}}$) limit (R. A. Scalzo et al. 2010; J. M. Silverman et al. 2011; S. Taubenberger et al. 2011). Because they are outliers in the Lira–Phillips relation, studies of their observed properties provide additional clues to the progenitors and explosion physics of SNe Ia. These "super-$M_{\mathrm{Ch}}$" SNe are initially supposed to come from the thermonuclear explosion of a super-$M_{\mathrm{Ch}}$ WD (D. A. Howell et al. 2006), for which gravitational collapse is resisted by the centrifugal force of a rapidly spinning WD (N. Langer et al. 2000; S. C. Yoon & N. Langer 2005). Such a super-$M_{\mathrm{Ch}}$ WD could come from a merger of two WDs (S. C. Yoon et al. 2007). However, as mentioned by S. Taubenberger (2017), simulations of the explosion of super-$M_{\mathrm{Ch}}$ WDs tend to give higher ejecta velocities (detonation; J. M. M. Pfannes et al. 2010a) or a lower mass of iron-group elements (deflagration; J. M. M. Pfannes et al. 2010b) than the observations. In conflict with the deflagration model, moreover, no carbon emission was reported in late-time spectra of any 03fg-like objects. Nevertheless, the merger of two WDs is still a popular channel to explain 03fg-like objects such as SN 2020esm (G. Dimitriadis et al. 2022) and SN 2021zny (G. Dimitriadis et al. 2023), as the luminous emission in ultraviolet bands (W. B. Hoogendam et al. 2024) and the long-lasting carbon absorption characteristics suggest an interaction between the SN ejecta and dense C/O-rich circumstellar material (CSM) that could be formed during the merger. The explosion within the CSM is also consistent with the core-degenerate scenario (C. Ashall et al. 2021 and references therein), where the merger of a WD and the degenerate core of an asymptotic giant branch star leads to an SN Ia within the C/O common envelope (CE), different from the double-WD merger scenario that requires the two WDs survive the CE phase.

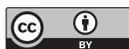

SN 2022pul, another 03fg-like object, was proposed to arise from a violent merger of two WDs based mainly on its centrally peaked [O I] and [Ne II] lines at ∼+300 days after peak brightness (L. A. Kwok et al. 2024; M. R. Siebert et al. 2024). Moreover, the burning of the secondary WD in a violent merger can explain the stronger [Ca II] lines of SN 2022pul compared with other 03fg-like objects (L. A. Kwok et al. 2024). In this Letter, we present late-time photometric and spectroscopic observations of SN 2022pul. We clearly identified the presence of prominent [C I] emission in the nebular spectrum of SN 2022pul, which has never been detected in any spectra of previous 03fg-like SNe Ia and has not been predicted by violent merger models.

## 2. Observations and Data Reduction

### 2.1. Photometry

We collected $BVgri$-band photometry of SN 2022pul, covering phases from $t \sim +140$ to 310 days after $B$-band maximum light, with the 0.8 m Tsinghua University-NAOC telescope (TNT; X. Wang et al. 2008; F. Huang et al. 2012) at the Xinglong Observatory of NAOC. Very late-time $gr$-band photometry was also obtained at $t \sim +512$ days with the Lijiang 2.4 m telescope (LJT; Y.-F. Fan et al. 2015) of Yunnan Observatories.

All these images were preprocessed following standard routines, including bias subtraction, flat-field correction, and cosmic-ray removal. Then, we applied point-spread function photometry with the Automated Photometry of Transients pipeline (AutoPhOT; S. J. Brennan & M. Fraser 2022) for both SN 2022pul and the local reference stars. No template subtraction is needed since the SN is far from its host galaxy and has a clean background. The instrumental magnitudes were calibrated against Gaia synthetic photometry (Gaia Collaboration et al. 2023; $BV$) and Pan-STARRS photometry release 1 (K. C. Chambers et al. 2016; $gri$).

### 2.2. Spectroscopy

Four late-time spectra were collected of SN 2022pul, including three (at $t \sim +142, +164, +194$ days relative to the $B$-band maximum) obtained with the BFOSC mounted on the Xinglong 2.16 m telescope (XLT; Z. Fan et al. 2016) and a very late spectrum taken at $t \approx +515$ days with the Low-resolution Imaging Spectrometer (LRIS; J. B. Oke et al. 1995) mounted on the 10 m Keck I telescope on Maunakea. The LRIS spectrum was reduced using the LPipe pipeline (D. A. Perley 2019), while standard IRAF routines were used to reduce the XLT spectra. Spectrophotometric standard stars were observed on the same nights to flux calibrate the SN spectra. We corrected for atmospheric extinction using the extinction curves of local observatories, and telluric lines were removed from the spectra. LRIS has an atmospheric-disperson corrector to minimize differential slit losses (A. V. Filippenko 1982).

## 3. Analysis and Discussion

### 3.1. Evolution of Light-curve Decline Rates

Our $BVgri$-band light curves are presented in Figure 1, together with the Zwicky Transient Facility (ZTF) data provided by Lasair,[7] the All-Sky Automated Survey for Supernovae (B. J. Shappee et al. 2014), the Las Cumbres Observatory and AAVSO data published by M. R. Siebert et al. (2024), and light curves of the typical normal SN 2011fe (R. E. Firth et al. 2015; K. Zhang et al. 2016; M. L. Graham et al. 2017) and another 03fg-like object SN 2009dc (S. Taubenberger et al. 2011; J. M. Silverman et al. 2012; B. E. Stahl et al. 2019). Following M. R. Siebert et al. (2024), we adopt the $B$-band maximum time and the explosion time as MJD 59808.3 ± 3.8 and MJD 59786.3, respectively. With the multiband light curves collected within ∼+140 to ∼+310 days after $B$-band maximum, we can identify an accelerated decline in the $B$ band after $t \approx +220$ days. In particular, the decline rates measured from linear fits to the TNT $B$-band data before and after $t \approx +220$ days are 1.63 ± 0.11 and 2.27 ± 0.43 mag per 100 days, respectively. The $g$-band light curve shows a similar accelerating decline after $t \approx +220$ days. The decline rate in the $V$ band remains around 2.01 mag per 100 days, and the evolution of the $r$-band light curve even slightly slows down. This enhanced fading in bluer bands is also observed in other 03fg-like objects such as SN 2009dc (S. Taubenberger et al. 2011) and has been attributed to dust formation and/or an unexpectedly early infrared (IR) catastrophe (S. Taubenberger et al. 2011). In comparison, the decline-rate changes in $Bg$-band light curves of SN 2011fe are much smaller. Note that the decline rates of SN 2022pul slow down later according to the $gr$-band data at $t \approx +512$ days, which is analogous to the evolution of SN 2009dc (S. Taubenberger et al. 2011).

### 3.2. Late-time Spectra

#### 3.2.1. Spectral Evolution

We show seven late-time spectra of SN 2022pul in Figure 2, including four spectra presented in this work ($t \approx +142, +163, +193, +515$ days) and three spectra taken from M. R. Siebert et al. (2024; $t \approx +221, +274, +335$ days). As reported by M. R. Siebert et al. (2024), clear emission lines of [O I] $\lambda\lambda 6300, 6364$ and [Ca II] $\lambda\lambda 7291, 7323$ can be seen in the late-time spectra of SN 2022pul. In the first three spectra, the strong emission feature in the 8600 Å region could be due to blends of Ca II IR and [Fe II] lines. We notice the appearance of a weak emission feature at around 8730 Å since $t \approx +220$ days, which was not discussed by M. R. Siebert et al. (2024). By $t \approx +515$ days, a strong emission feature appears around 9850 Å. According to the wavelength of the emission peak, these two features could be attributed to [C I] $\lambda 8727$ and [C I] $\lambda\lambda 9824, 9850$, respectively; note that these carbon lines have been used to explain the late-time spectra of the Type Ic SN 2007gf (P. A. Mazzali et al. 2010). Unlike the centrally peaked [O I] and [C I] lines, the velocity of the [Ca II] $\lambda\lambda 7291, 7321$ lines is initially blueshifted and then gradually decreases. In this work, we find flat-topped and centrally peaked [Ca II] $\lambda\lambda 7291, 7321$ features in the $t \approx +515$ days spectrum.

---

[7] https://lasair-ztf.lsst.ac.uk

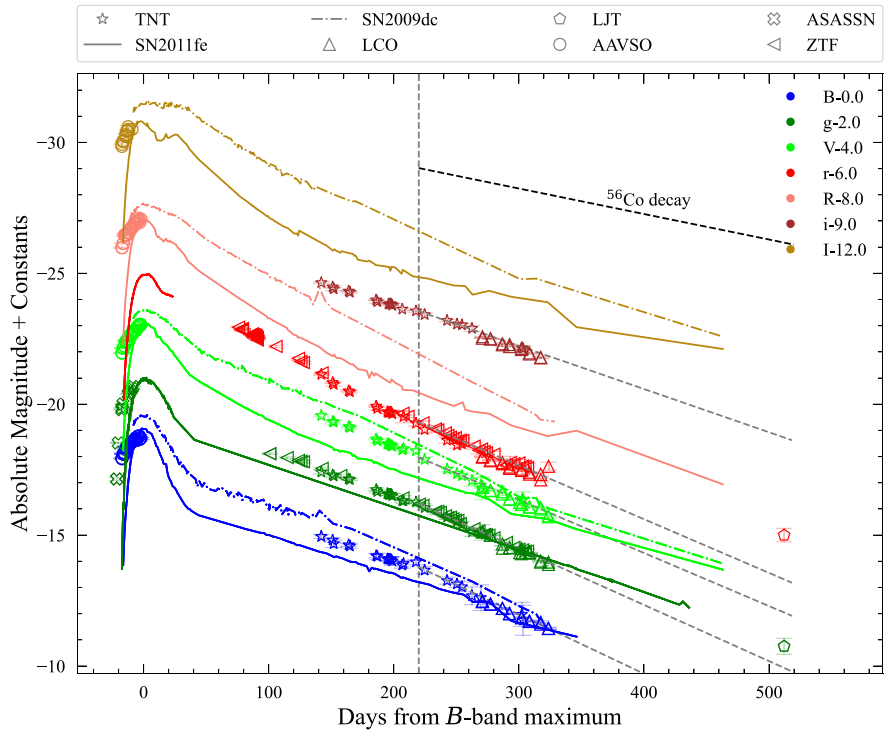

**Figure 1.** Optical light curves of SN 2022pul obtained at phases from $t \approx +140$ to $+520$ days relative to the $B$-band maximum. Light curves of different filters are shifted vertically for clarity. The gray vertical dashed line denotes the epoch after which the $B$-band light curve shows a faster decline rate. The gray dashed lines represent the best linear fits to different bands of SN 2022pul within $\sim+220$–310 days, while the black dashed line shows the decline rate of $^{56}$Co decay. The solid lines and dash-dotted lines represent the light curves of the well-observed normal type Ia supernova SN 2011fe (K. Zhang et al. 2016) and super-$M_{ch}$ object SN 2009dc.

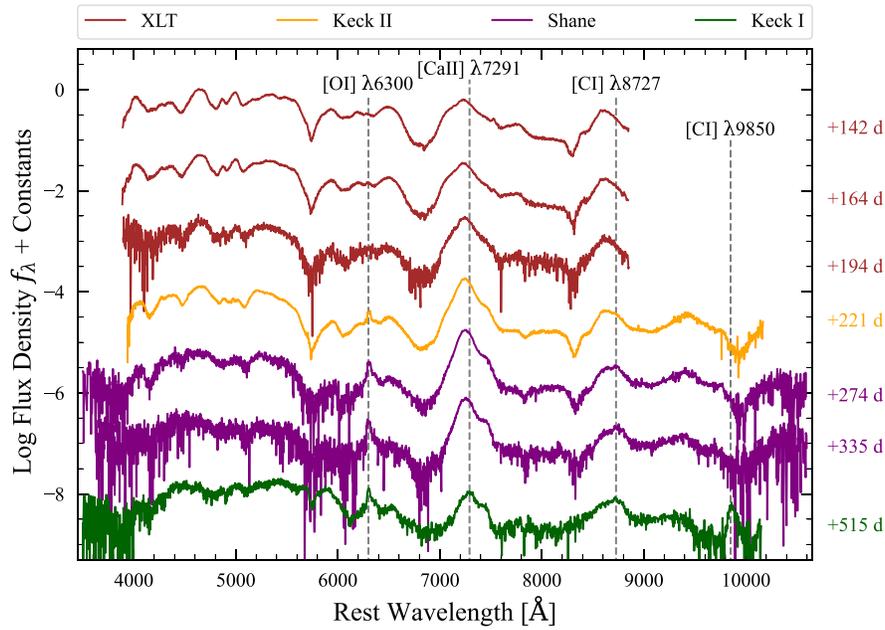

**Figure 2.** Spectral evolution of SN 2022pul, spanning the phase from $t \approx +142$ to $+515$ days after maximum light. All spectra have been corrected for Galactic reddening and host redshift. Spectra taken with different telescopes are marked in different colors. The spectral epoch is shown on the right side of each spectrum. The rest-frame wavelengths of [O I] $\lambda6300$, [Ca II] $\lambda7291$, [C I] $\lambda8727$, and [C I] $\lambda9850$ are labeled by gray dashed lines.

### 3.2.2. Distributions of Elements

To examine the distribution of different elements in the SN ejecta, several lines of the spectra taken at $t \approx +274$ and $+515$ days are shown in velocity space in Figure 3. The conversion from wavelength to velocity is based on the rest-frame wavelength of the corresponding line. Lines of iron-group elements are not examined here because of their serious blending. At $t \approx +274$ days, the [C I] $\lambda8727$ and [O I] $\lambda6300$ lines show a slightly redshifted center and a narrow full width at half-maximum intensity of $\sim2000$ km s$^{-1}$, in contrast with the wide and blueshifted [Ca II] $\lambda7291$ line. In addition, the [C I] $\lambda8727$ and [O I] $\lambda6300$ lines show a flat-topped shape with a width of $\sim1400$ km s$^{-1}$. Owing to the homologous expansion of the ejecta, the similarity of the line profiles of [O I] $\lambda 6300$ and [C I] $\lambda 8727$ indicates a similar spatial distribution of carbon and oxygen or at least a similar emission region.

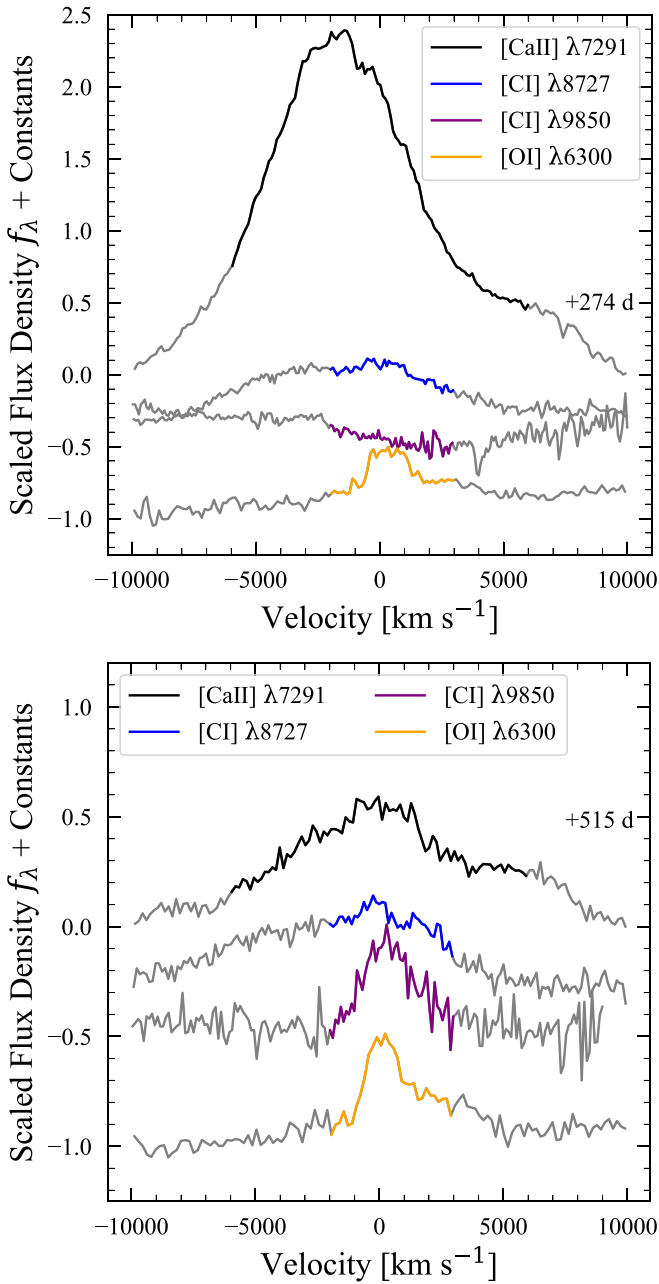

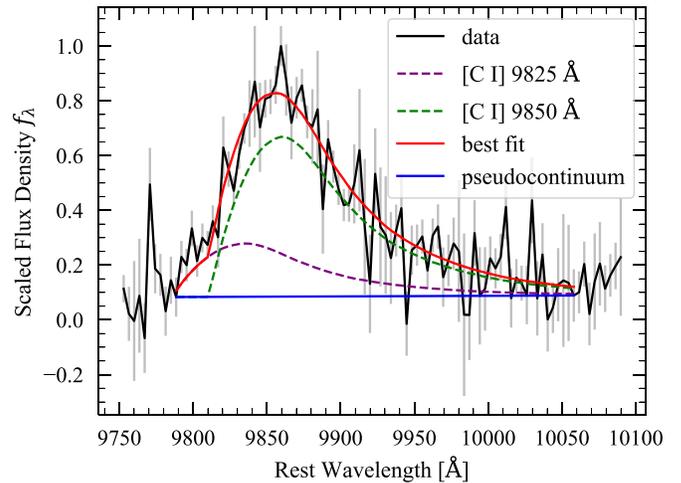

**Figure 3.** Velocity distributions of [C I], [O I], and [Ca I] features at $t \approx +274$ (top panel) and $t \approx +515$ days (bottom panel). The regions that may be contributed by features are roughly chosen and highlighted with corresponding colors. The flux density of different features of the same spectrum is scaled to an arbitrary value.

Although the [C I] $\lambda 9850$ feature is not visible in the spectrum taken at $t \approx +274$ days, it appears noticeable in the $t \approx +515$ days spectrum, with a velocity distribution similar to that of the [O I] $\lambda 6300$ line. Note that the [C I] $\lambda 9850$ feature is characterized by an asymmetric profile with an extended red wing, but this might not indicate an asymmetric distribution of carbon. Note that strong thermal dust emission was found in SN 2022pul (M. R. Siebert et al. 2024); thus, such an asymmetric profile could be fit by considering a uniform sphere and isotropic dust scattering (see, e.g., A. Jerkstrand 2017), and details of the calculation can be found in Appendix B. As shown in Figure 4, we obtained a good match with a scattering optical depth of $\tau = 1.9$, a velocity shift of the emission sphere of $v_{\rm shift} = 682^{+54}_{-49}$ km s$^{-1}$, and a maximum velocity of the emission region of $v_{\rm max} = 1898^{+94}_{-85}$ km s$^{-1}$. Note that the [O I] $\lambda 6300$ line also shows a similarly asymmetric profile, which, however, may suffer from blending with the [O I] $\lambda 6364$ line in the red wing. The line profile of [C I] $\lambda 8727$ is very uncertain owing to relatively large noise and blending with other lines by $t \approx +515$ days.

**Figure 4.** Dust scattering fits to the [C I] $\lambda\lambda 9824$, 9850 lines at $t \approx +515$ days. The spectrum, after corrections for host-galaxy redshift and Milky Way extinction, is shown in black and the uncertainty is in gray. The blue line represents the pseudocontinuum defined as a straight line connecting the red and blue sides of the features.

The blueshifted velocity of [Ca II] $\lambda 7291$ at $t \approx +274$ days indicates an off-center distribution of calcium (L. A. Kwok et al. 2024). At $t \approx +515$ days, however, we find that [Ca II] has a flat-topped shape and a near-zero velocity shift. The asymmetric profile of [Ca II] might be due to a blend with other lines such as [Fe II] and [Ni II] instead of an asymmetric distribution; otherwise, the relative flux of the peak to the red wing should change little with time, which is inconsistent with the observations.

### 3.3. Modeling the [C I] $\lambda\lambda 9824$, 9850 Emission Features

L. A. Kwok et al. (2024) ran the violent merger model from R. Pakmor et al. (2012) with CMFGEN (D. J. Hillier & L. Dessart 2012) to $\sim$316 days and found that the [Fe III] lines are too strong and the [O I] $\lambda\lambda 6300$, 6364 lines are too weak compared with the observations. Then they added clumping of the ejecta to reduce the ionization stages and added mass in the innermost region to account for [O I]. The result of this clumped version of the merger model matches the observed spectrum well.

We tried to model the [C I] $\lambda\lambda 9824$, 9850 lines in our Keck spectrum with the clumped merger model of L. A. Kwok et al. (2024) calculated for a phase at $t \approx +514$ days (corresponding to about 536 days after the explosion) using CMFGEN. As shown in Figure 5, however, the synthetic spectrum does not exhibit any significant signature of C I emission, as little carbon is left in the central region in this model. The flux density of the synthetic spectrum is also higher than that observed.

To improve the fitting, we revised the model by setting the carbon mass as that of the oxygen ($\sim 0.06 M_\odot$) in the central region (where the expansion velocity is $\lesssim 2000$ km s$^{-1}$). We also made the model have more clumpy ejecta with a filling factor of 0.2 (in contrast to the original value of 0.5), which can decrease the ionization and excitation of the gas and suppress the emission flux of ions. The result of the revised model shows a clear signature of [C I] $\lambda\lambda 9824, 9850$ lines, with a flux density comparable to the observations (see Figure 5). We note that the [C I] lines are hardly identified owing to their blending with other lines if the clumpiness is not enhanced, which could account for the "absence" of [C I] $\lambda\lambda 9824, 9850$ lines at earlier times.

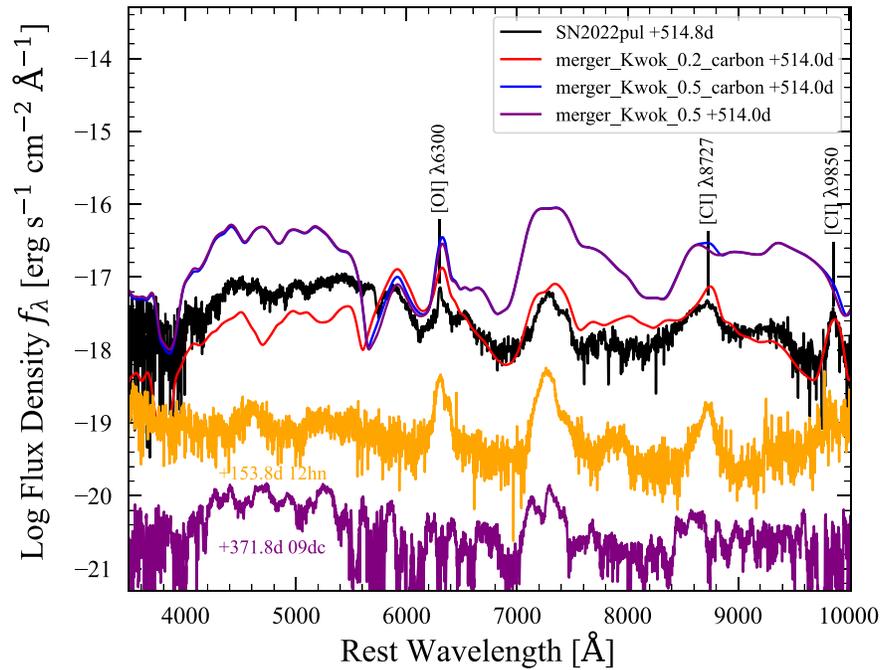

**Figure 5.** Model spectra compared to that of SN 2022pul obtained at $t \approx +514.8$ days. Late-time spectra of SN 2009dc and SN 2012hn are also presented for comparison. The observed spectra have been corrected for host-galaxy redshift and Milky Way extinction. [O I] and [C I] lines are labeled in the plot. The original model of L. A. Kwok et al. (2024) is labeled as "merger_Kwok_0.5 +514.0d," and our revised model is labeled as "merger_Kwok_0.5_carbon +514.0d" (including carbon) and "merger_Kwok_0.2_carbon +514.0d" (including carbon and enhancing clumpiness). The spectra of SN 2009dc and SN 2012hn are shifted vertically for clarity.

In Section 3.2.2, we used the dust scattering mechanism to interpret the asymmetric line profiles of [C I]. However, the temperature in the central region of our model is ∼3000 K, which is higher than the dust condensation temperature (as high as 2000 K; A. G. G. M. Tielens et al. 2005) and thus disfavors dust formation in the innermost region. However, this could be due to our rough model aiming mainly to confirm the origin of 9850 Å emission features from [C I] $\lambda\lambda 9824, 9850$ lines. More accurate temperatures as well as other ejecta properties (e.g., the mass ratio of carbon to oxygen) could be estimated by more precise models in future work.

### 3.4. Origin of Carbon

In this work, carbon lines with a profile similar to that of [O I] are found in late-time spectra of a 03fg-like SN Ia for the first time. This is an important clue to explore the explosion mechanisms of these peculiar objects, since viable models should reproduce such rare line features in late-time spectra.

The lines of centrally peaked, narrow [C I] and [O I] in SN 2022pul, together with the strong [Ca II] lines, resemble Ca-rich objects such as SN 2012hn (S. Valenti et al. 2014; see Figure 5). A possible explosion model of Ca-rich objects is a merger of a CO WD and a hybrid He–CO WD, in which a weak He detonation is triggered in the surface of the He–CO WD when accreting the CO material, while the core of the He–CO WD is left behind (Y. Zenati et al. 2023). This avenue is similar to the violent merger model that is suggested to match best with the observations of SN 2022pul (L. A. Kwok et al. 2024; M. R. Siebert et al. 2024), but the difference is that the latter one ignites a carbon detonation in a more massive primary WD, which can lead to the much higher peak luminosity of SN 2022pul ($M_B \approx -18.9$ mag; M. R. Siebert et al. 2024) than that of SN 2012hn ($M_B \approx -15.1$ mag; S. Valenti et al. 2014). Note that the composition of the central region of the violent merger model is purely oxygen that comes from the burning of the companion (R. Pakmor 2017), in conflict with the significant carbon of SN 2022pul in the center. Perhaps the helium shell of the secondary WD is also ignited when hit by the shock wave of the explosion of the primary WD (R. Pakmor et al. 2022), and then the secondary WD explodes as a carbon-rich SN so that there exists carbon in the innermost region. The viability of this avenue needs to be examined in future work.

Pure deflagration models, which can naturally explain the [C I] emission, are also viable for SN 2022pul. In three-dimensional simulations of deflagration, a significant amount of carbon and oxygen mix into the central region of the ejecta owing to the downdrafts of unburnt matter between rising burning plumes (C. Kozma et al. 2005; M. Fink et al. 2014). Although the near-$M_{Ch}$ pure deflagration models (M. Fink et al. 2014) could be disfavored due to the low masses of synthetic Ni$^{56}$ and intermediate-mass elements (L. A. Kwok et al. 2024), the super-$M_{Ch}$ models (J. M. M. Pfannes et al. 2010b) could solve these problems. In addition, SN 2022pul was supposed to explode within dense C/O-rich CSM that favors an explosion occuring after the disruption of the secondary WD (M. R. Siebert et al. 2024), which is consistent with the pure deflagration model. For the violent merger model, the CSM might not be massive enough since the explosion occurs before the complete disruption of the secondary WD, for which a third star is invoked (L. A. Kwok et al. 2024). Hence, the pure deflagration of a super-$M_{Ch}$ WD seems to be a more natural model for SN 2022pul. In either case, the mass ratio of carbon to oxygen in the innermost region of the ejecta should not be lower than that in the progenitor WD since part of the oxygen could be produced by burning carbon during the explosion. Therefore, if the C/O ratio of ∼1 adopted in our model (see Section 3.3) is correct, the C/O ratio of the progenitor should be $\gtrsim 1$.

Although the pure deflagration of a super-$M_{Ch}$ WD is an attractive model to explain SN 2022pul, it cannot account for the bulk of 03fg-like objects owing to the rare detection of [O I] or [C I] emission features in these SNe. For most 03fg-like SNe Ia, a detonation mechanism is preferred to account for a high $^{56}$Ni mass characteristic of them (e.g., SN 2009dc). Note that SN 2022pul is the third 03fg-like object that displays convincing [O I] emission (M. R. Siebert et al. 2024).

## 4. Conclusion

In this Letter, we present new photometric and spectroscopic observations of SN 2022pul at late times. Our main results are as follows:

(i) Similar to other 03fg-like objects, SN 2022ful shows an enhanced flux decline in bluer bands after $t \approx +220$ days since the time of peak brightness, and the decline rates of its light curves slow down before $t \approx +500$ days.

(ii) Differently from the blueshifted velocities at earlier times, the [Ca II] lines have a near-zero velocity shift and a flat-topped profile at $t \approx +515$ days, suggesting that the calcium distribution is not as asymmetric as previously suggested.

(iii) Centrally peaked [C I] emission lines are found to have a profile similar to that of the [O I] lines. This indicates that carbon and oxygen are located in the innermost regions of the SN ejecta, which could result from explosion of the secondary WD as a Ca-rich SN during the merger of two WDs, while a more natural explanation could be due to pure deflagration of a super-$M_{Ch}$ WD.

(iv) The asymmetric profile of the [C I] $\lambda\lambda 9824, 9850$ lines can be well reproduced by dust scattering, but whether the dust can be formed in the innermost region of the ejecta needs further exploration.

Very late-time spectra at $t \gtrsim +500$ days can provide important clues, such as the prominent [C I] $\lambda\lambda 9824, 9850$ lines, to study the explosion mechanisms of 03fg-like objects. In addition, they allow a method of constraining the mass ratio of carbon to oxygen in the progenitor WD. Spectra should cover the near-IR region to detect [C I] $\lambda\lambda 9824, 9850$. Studies of the viability of inducing a Ca-rich SN in a violent merger and more detailed comparisons with super-$M_{Ch}$ deflagration models are needed to further constrain the explosion mechanism of SN 2022pul.

### Acknowledgments

This work is supported by the National Natural Science Foundation of China (NSFC grants 12288102, 12033003, and 11633002), the science research grant from the China Manned Space Project No. CMS-CSST-2021-A12, and the Tencent Xplorer Prize. A.V.F.'s research group at UC Berkeley received financial assistance from the Christopher R. Redlich Fund, as well as donations from Gary and Cynthia Bengier, Clark and Sharon Winslow, Alan Eustace, William Draper, Timothy and Melissa Draper, Briggs and Kathleen Wood, and Sanford Robertson (W.Z. is a Bengier-Winslow-Eustace Specialist in Astronomy, T.G.B. is a Draper-Wood-Robertson Specialist in Astronomy, and Y.Y. was a Bengier-Winslow-Robertson Fellow in Astronomy), as well as numerous other donors. J.-J. Zhang is supported by the International Centre of Supernovae, Yunnan Key Laboratory (No. 202302AN360001), the National Key R&D Program of China (No. 2021YFA1600404), NSFC grant 12173082, the Yunnan Province Foundation (grant 202201AT070069), the Top-notch Young Talents Program of Yunnan Province, and the Light of West China Program provided by the Chinese Academy of Sciences.

We acknowledge Stéphane Blondin for his help with the setup of CMFGEN. We thank the staff at the various observatories where data were obtained. Some of the data presented herein were obtained at the W. M. Keck Observatory, which is operated as a scientific partnership among the California Institute of Technology, the University of California, and NASA; the observatory was made possible by the generous financial support of the W. M. Keck Foundation. This work was partially supported by the Open Project Program of the Key Laboratory of Optical Astronomy, National Astronomical Observatories, Chinese Academy of Sciences.

*Facilities:* Keck:I (LRIS), YAO:2.4m, Beijing:2.16m, and PO:1.2m (ZTF).

*Software:* AutoPhOT (S. J. Brennan & M. Fraser 2022), CMFGEN (D. J. Hillier & L. Dessart 2012), emcee (D. Foreman-Mackey et al. 2013), Matplotlib (J. D. Hunter 2007), NumPy (C. R. Harris et al. 2020), SciPy (P. Virtanen et al. 2020).

## Appendix A
## Photometric and Spectroscopic Observations

The photometric data and the journal of spectroscopic observations are presented in Tables 1 and 2, respectively. The spectra will be available in the Weizmann Interactive Supernova Data Repository.[8]

---

[8] https://www.wiserep.org/

**Table 1**
Observed Photometry of SN 2022pul

| MJD | Magnitude | Error[a] | Band | Telescope |
|---|---|---|---|---|
| 59950.778 | 16.073 | 0.075 | B | TNT |
| 59950.781 | 15.453 | 0.048 | V | TNT |
| 59950.785 | 15.614 | 0.072 | g | TNT |
| 59950.788 | 15.884 | 0.036 | r | TNT |
| 59950.792 | 15.377 | 0.033 | i | TNT |
| 59959.899 | 16.210 | 0.079 | B | TNT |
| 59959.903 | 15.682 | 0.042 | V | TNT |
| 59959.906 | 15.720 | 0.061 | g | TNT |
| 59959.910 | 16.193 | 0.036 | r | TNT |
| 59959.914 | 15.542 | 0.046 | i | TNT |
| 59960.764 | 16.351 | 0.142 | B | TNT |
| ... | ... | ... | ... | ... |
| 60112.563 | 18.698 | 0.218 | g | TNT |
| 60112.577 | 17.896 | 0.101 | i | TNT |
| 60321.896 | 22.267 | 0.307 | g | LJT |
| 60321.904 | 22.020 | 0.257 | r | LJT |

**Note.**
[a] $1\sigma$.

(This table is available in its entirety in machine-readable form in the online article.)

**Table 2**
Overview of Optical Spectra of SN 2022pul

| MJD | Date | Phase[a] | Range (Å) | Exposure (s) | Instrument/Telescope |
|---|---|---|---|---|---|
| 59950.9 | 20230106 | 142.1 | 3903–8876 | 3000 | BFOSC/XLT |
| 59972.8 | 20230128 | 164.0 | 3899–8871 | 3300 | BFOSC/XLT |
| 60002.8 | 20230227 | 193.9 | 3907–8877 | 3300 | BFOSC/XLT |
| 60324.6 | 20240115 | 514.8 | 3137–10180 | 2894 | LRIS/Keck I |

**Note.**
[a] Relative to the B-band maximum light, $MJD_{Bmax} = 59808.3$.

(This table is available in machine-readable form in the online article.)

## Appendix B
## Fitting of Asymmetric Line Profile

The asymmetric line profile with a red extended tail could be due to isotropic dust scattering (see, e.g., A. Jerkstrand 2017), since the comoving frame wavelength is always lower than in the original emitting frame. Assuming a uniform emission coefficient and scattering probability in a sphere, we used a Monte Carlo method to compute the scattering line profile of optical depths $\tau$ (in the center) from 0.1 to 5.0 with an interval of 0.1. Specifically, we randomly generated 5 million photons in the sphere and simulated their scattering to obtain the final wavelength distribution. Each photon has a random initial velocity, travels in the sphere, changes its wavelength if scattered, and gets out of the sphere with the final wavelength. In Figure 6, we show four profiles of $\tau = 0, 1, 2, 3$, where the unit of wavelength is $\Delta\lambda/\lambda_0/(v_{max}/c)$, $\Delta\lambda$ is the deviation from the rest wavelength $\lambda_0$ of the line, $v_{max}$ is the maximum velocity of the emission region, and $c$ is the speed of light. For a given scattering profile, we used three parameters – strength scale factor, velocity shift (the bulk motion of the sphere along the line of sight), and $v_{max}$ – to fit the observation. Then we fitted the observed line using each scattering profile to find the best $\tau$ with the Python package `scipy.curve_fit`. Under the best $\tau$, we used the Markov Chain Monte Carlo Hammer `emcee` (D. Foreman-Mackey et al. 2013) to calculate the uncertainties of the parameters.

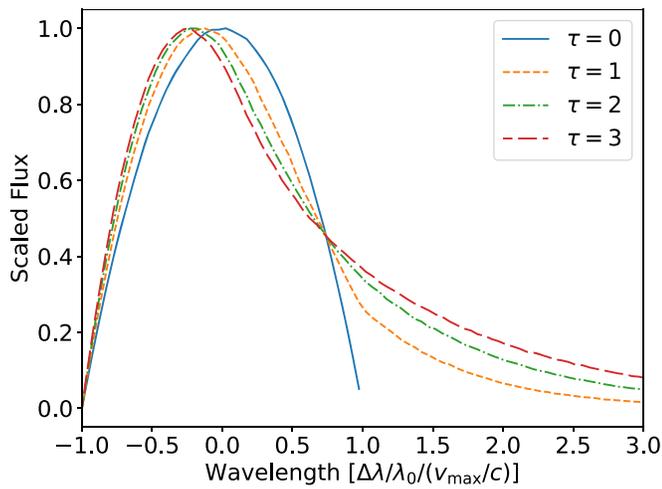

**Figure 6.** Line profiles from dust scattering opacity in a uniform sphere. The profiles have been scaled according to the peaks.

## ORCID iDs

Jialian Liu ⓘ https://orcid.org/0009-0000-0314-6273
Xiaofeng Wang ⓘ https://orcid.org/0000-0002-7334-2357
Alexei V. Filippenko ⓘ https://orcid.org/0000-0003-3460-0103
Thomas G. Brink ⓘ https://orcid.org/0000-0001-5955-2502
WeiKang Zheng ⓘ https://orcid.org/0000-0002-2636-6508
Jujia Zhang ⓘ https://orcid.org/0000-0002-8296-2590
Shengyu Yan ⓘ https://orcid.org/0009-0004-4256-1209